# A caloritronics-based Mott neuristor

Javier del Valle*, Pavel Salev, Yoav Kalcheim and Ivan K. Schuller

Department of Physics and Center for Advanced Nanoscience, University of California-San Diego, La Jolla, California 92093, USA

*Corresponding author: jdelvallegranda@ucsd.edu

**Machine learning imitates the basic features of biological neural networks to efficiently perform tasks such as pattern recognition[1,2]. This has been mostly achieved at a software level, and a strong effort is currently being made to mimic neurons and synapses with hardware components, an approach known as neuromorphic computing[3–6]. CMOS-based circuits have been used for this purpose[4–7], but they are non-scalable, limiting the device density and motivating the search for "neuromorphic materials"[8]. While recent advances in resistive switching have provided a path to emulate synapses at the 10 nm scale[9–13], a scalable neuron analogue is yet to be found. Here, we show how heat transfer can be utilized to mimic neuron functionalities in Mott nanodevices. We use the Joule heating created by current spikes to trigger the insulator-to-metal transition in a biased $VO_2$ nanogap. We show that thermal dynamics allow the implementation of the basic neuron functionalities: activity, leaky integrate-and-fire, volatility and rate coding[7,14]. By using local temperature as the internal variable, we avoid the need of external capacitors, which reduces neuristor size by several orders of magnitude. This approach could enable neuromorphic hardware to take full advantage of the rapid advances in memristive synapses, allowing for much denser and complex neural networks. More generally, we show that heat dissipation is not always an undesirable effect: it can perform computing tasks if properly engineered.**

Machine learning has experienced an unprecedented growth in recent years, often referred to as an "artificial intelligence revolution"[1,2]. Its fundamental approach is inspired by biological systems: using neural networks to classify large amounts of data into sorting categories. Classic examples are speech and image recognition [1,2]. Neural networks are composed of two basic elements: neurons and synapses. Current machine learning schemes implement these elements at a software level: neurons and synapses are simulated on standard computers based on a von Neumann architecture[1,2]. This approach is inefficient in terms of computation speed and energy consumption, motivating a great effort towards developing hardware-based neuromorphic systems[3,5]. CMOS-based circuitry has been successfully used for this, allowing to build tuneable and energy efficient neural networks[4,6,7]. Unfortunately, CMOS-based components rely on combinations of multiple transistors and capacitors that make them complex and large[7]. This limits circuit scalability and, hence, poses a limitation to achieve dense neural networks which could eventually rival the brain.

A solution to this problem might be found in "neuromorphic materials", whose intrinsic properties mimic those of neurons and synapses[8,15]. Resistive switching (RS), a phenomenon in which an applied electric field modifies the resistance of the material, offers a unique opportunity to achieve this goal[13,16,17]. RS can be volatile[18–20] or non-volatile[9,21], giving the possibility to emulate synapse or neuron behaviours, respectively. Multiple groups have used non-volatile RS to achieve synaptic functionalities[11,22], and memristor crossbar arrays have already been used to perform pattern recognition[10,12,23]. These synapse realizations, however, still rely on traditional electronics to play the role of neurons (neuristors). This approach does not take full advantage of the scalability and simplicity offered by memristive synapses, and motivates the search for a simpler and more scalable neuristor.

A neuristor must feature the most basic functionalities of real neurons: i) activity (outputting a current), ii) leaky integrate-and-fire, iii) volatility (resetting after a firing), and iv) rate coding of the external stimuli. Several groups have explored the possibility of a RS-based neuristor. Pickett et al.[24] and Ignatov et al.[25] use volatile RS in Mott insulators to emulate neurons, but their solution relies on large capacitors and is non-scalable. Wang et al.[26], Tuma et al.[27] and Stoliar et al.[28] successfully implement integrate and fire dynamics using diffusive memristors, phase-change materials and Mott insulators, respectively. However, these systems are not active as they do not generate a current. Moreover, they are not volatile i.e. do not reset automatically, a characteristic needed for spiking dynamics. This limits their autonomy and their practical implementation as standalone neurons. A fully autonomous and scalable neuristor is still to be found.

Performing leaky integrate-and-fire is one of the key functionalities of a neuristor. It must integrate all the stimuli coming from previous neurons and fire when the excitation is above a certain threshold[14]. This process can be viewed as two separate ones: the leaky integration, and the firing mechanism (Figure 1a). In the case of biological neurons, the cell membrane acts as a capacitor that integrates incoming ionic currents. The firing mechanism, based on voltage gated sodium and potassium channels, is activated once the membrane potential passes a certain threshold. CMOS neuristors use a similar approach: a capacitor plays the role of cell membrane by integrating current from incoming pulses[7], while the CMOS circuitry produces the firing events. While this parallelism with biological systems is appealing, the use of capacitors to store the internal state of the neuron limits the circuit scalability. In order to avoid malfunctioning, their capacitance must be much larger than the parasitic capacitance of the electrode lines. Therefore, integration capacitors cover a large area in current CMOS neurons, having lateral sizes in the order of 10-100 μm[7]. Capacitor downscaling is one of the most challenging issues in other technologies such as DRAM, where the industry has dedicated intense effort towards developing complex 3D capacitive structures to circumvent this problem[29].

Instead of electrical currents, we propose using heat flow to perform computing tasks, an approach known as caloritronics. Temperature substitutes charge as the integrating variable, as depicted in Figure 1b. Current spikes coming from previous neurons induce

Joule heating while passing through a resistive element (heater), increasing local temperature with every spike. The heater is thermally coupled to a firing element that is very sensitive to temperature changes and fires once a threshold temperature is exceeded. In this work, we use $VO_2$, a well-known correlated oxide with a sharp insulator-to-metal transition (IMT) around 340 K[30], as the firing element. We realize leaky integrate-and-fire using the thermal dynamics, which are governed by similar equations to those describing the charge dynamics of a leaky capacitor (see figures 1a and 1b). Adopting local temperature as the internal state allows building simple-design neuristors that can be downsized to the nanoscale, as thermal dynamics equations preserve the same form independent of the system size.

We fabricated and tested a proof-of-concept neuristor that performs all basic neuronal functionalities. It consists of a $VO_2$ thin film on top of which two layers of electrodes are patterned (detailed fabrication process in methods section and Extended Data Figure 1). The first layer consists of two Ti/Au electrodes (running vertically in Figure 2a) separated by a 50 nm gap. These electrodes are used to apply voltage to the $VO_2$, and provide the source of the neuristor's active output. If the voltage is high enough, a transition into the metallic phase can be electrically triggered[18,19,31]. Figure 2b shows the current as a function of time when a voltage pulse is applied to the gap. It illustrates the threshold nature of the voltage triggered IMT: the device becomes metallic once a threshold voltage ($V_{Th}$) is exceeded[31]. The second electrode layer is a Ti/Au nanowire (running horizontally in Figure 2a) which acts as a heater. It is separated from the bottom electrodes by a 70-nm-thick $Al_2O_3$ layer which provides electrical insulation (resistance larger than 20 MΩ), but ensures thermal coupling between them.

Figure 2c shows $V_{Th}$ of the $VO_2$ gap as a function of temperature. Two cases are shown: 0 mA (black) and 12.5 mA (red) current flowing through the nanowire heater (horizontal electrode). In the second case, Joule heating locally increases the temperature of the gap reducing its $V_{Th}$. To work as a neuristor, the gap is kept under a DC bias just below its threshold voltage ($V_{DC}<V_{Th}$), as represented in Figure 2c by a green dot. When a high enough current pulse $I_{Input}$ is passed through the heater, it lowers $V_{Th}$ below $V_{DC}$, and the gap turns metallic, generating an output current through the bottom electrodes ($I_{Output}$). This situation is presented in Figure 2d, where a 30 ns current pulse is applied to the heater, triggering the IMT in the gap. We must emphasize that the output and input electrodes are electrically isolated, and an output current is generated when the neuristor fires, making it an active element. Our device releases energy when stimulated, a crucial property to avoid a re-amplification stage after each neural layer. This goes one step beyond previous RS-based neuristors[26–28], which become conducting after performing integrate-and-fire but do not create an output on their own.

Volatility is a necessary feature to implement spiking dynamics. As presented so far, our device would remain conductive once triggered. For it to reset after the firing event, we add a resistor ($R_{Load}$) in series with the gap (Inset figure 2e). The resistor value must be in between $R_{OFF}$ (20 kΩ) and $R_{ON}$ (100 Ω), the gap resistance in the insulating and metallic states, respectively[32]. The role of this resistor is to lower the voltage across the gap once the $VO_2$ becomes metallic[32], which in turn reduces heat dissipation and decreases local temperature after the firing event. As a result, the $VO_2$ returns to its insulating state, resetting the neuristor (Figure 2e). The result is the desired effect: spike in – spike out.

Leaky integrate-and-fire (LIF) dynamics are governed by the characteristic thermal times of our device, given by its specific heat and thermal resistance to the substrate. Figure 3a shows the warm up times of the neuristor as a function of $I_{Input}$, that is, how long it takes the device to warm and fire once the current flows through the heater. Typical times are on the order of 10-100 ns. Using pulse widths and rates around that timescale allow us to implement LIF dynamics. Figure 3b shows the response of the neuristor ($I_{Output}$) when a train of current pulses is sent to the heater ($I_{Input}$). The first pulse does not raise the temperature enough to fire the device, but the cumulative effect of several pulses adds up to trigger the IMT after an integration period. The number of pulses necessary to produce a firing event depends on the pulse amplitude. Figure 3c shows the probability of the neuristor firing after a certain number of pulses are applied to the input. Several current amplitudes are shown, offering a clear visualization of the LIF dynamics. For high current pulses, the device is always triggered with just one pulse. For lower currents, the probability of it firing with just one pulse goes down, and more pulses are necessary to induce the IMT. The overall mean integration time is shifted to larger values as the current amplitude is decreased. When the input current is too low, heat leakage into the environment overcomes the dissipated power, and the device does not fire for any number of pulses.

Another basic feature of biological neurons is rate coding: the frequency at which a neuron spikes depends on the amplitude of its stimulus. Strong stimulus will produce high frequency spiking, while weak stimulus will yield slower patterns[14]. Our neuristor reproduces that feature, as shown in figure 4a. A constant current is passed through the heater, resulting in a repetitive spiking output. The frequency of the output increases with the input current (figure 4b). After the neuristor fires, both the temperature and the voltage across the gap drop, leaving the system in a refractory period until they increase back to their initial values. The time it will take the temperature to return to its firing value is given by the warm up time shown in figure 3a. The device will heat up faster at higher input currents, leading to the desired rate coding. On the other hand, the recovery time for the voltage will be proportional to $R_{OFF} \cdot C_{Par}$, where $C_{Par}$ is the parasitic capacitance of the electrode lines. We must note that $R_{OFF}$ is not constant, but decreases with temperature (inset in figure 4b). This makes the charge up time faster at higher local temperatures (higher input currents), also leading to the desired rate coding. Whether the origin of the rate coding is given by heat dissipation or by the parasitic capacitance will depend on the particular geometry and characteristics of the device. In our particular case, the period between spikes is in the μs range, 10-20 times slower than typical warm up times. This suggest that the charging time is the more dominant factor determining the recovery time in our system. Further optimization, such as embedding the load resistor into the lithography, could reduce $C_{Par}$, to the point in which rate coding would be governed solely by thermal dynamics.

Many of the relevant parameters of the proposed neuristor depend on the particular device design, as well as on the intrinsic properties of the chosen materials. This gives plenty of room to explore and improve its functionalities. Different substrates, insulating spacers or geometric designs will strongly change the device properties. For instance, a less thermally conductive substrate would reduce heat leakage to the environment, allowing for the use of lower input currents. A similar effect could be achieved by further reducing sizes. A larger ON/OFF ratio between insulator and metal would result in higher output currents when the IMT is triggered. With proper optimization, the output current could become larger than the input current, resulting in signal amplification. This could allow to use the neuristor as a completely independent device, without the need of a dedicated amplification stage.

Caloritronics and resistive switching can be combined to create scalable and autonomous neuristors. We demonstrated four basic neural functionalities: activity, volatility, leaky integrate-and-fire dynamics and rate coding using simple devices that can be downscaled well below the μm scale. Combined with the fast advances in memristor technology, this could pave the way to develope dense neuromorphic hardware, allowing for deeper and more complex neural networks. Our approach could be generalized to other physical phenomena. Other systems at the edge of a phase transition very sensitive to external stimuli might show similar behaviour. On a broader scope shows that, although often regarded as an undesirable consequence, power dissipation might actually enable new ways of computing, taking advantage of the rich phenomenology of correlated systems.

## Acknowledgments

This work was supported as part of the Quantum Materials for Energy Efficient Neuromorphic Computing (Q-MEEN-C) Energy Frontier Research Center (EFRC), funded by the U.S. Department of Energy, Office of Science, Basic Energy Sciences under Award # DE-SC0019273. Part of the fabrication process was done at the San Diego Nanotechnology Infrastructure (SDNI) of UCSD, a member of the National Nanotechnology Coordinated Infrastructure (NNCI), which is supported by the National Science Foundation under grant ECCS-1542148. J. del Valle thanks Fundación Ramón Areces for the support with a postdoctoral fellowship. The authors thank Marcelo J. Rozenberg, Juan Trastoy and George Kassabian for helpful discussions.

## Author Contributions

J.d.V. and I.K.S. conceived the idea. J.d.V, and Y.K. designed and fabricated the devices. J.d.V. and P.S. performed the transport measurements and analyzed the data. J.d.V. and I.K.S. wrote the manuscript. All authors participated in the discussion of the results and corrected multiple iterations of the manuscript.

## Data availability

The data supporting the plots and claims of this manuscript are available from the corresponding authors upon reasonable request.

## Methods

**Sample preparation.** A 70 nm $VO_2$ film was grown by reactive sputtering on top of an R-cut $Al_2O_3$ substrate. A 4 mtorr Argon/Oxygen mix (8% $O_2$) was used during deposition. The substrate temperature was kept at 520°C, and cooled down after sputtering at a rate of 12°C/min. X-ray diffraction shows textured orientation along $\langle 100 \rangle$ for $VO_2$. Transport measurements show a four orders of magnitude IMT, confirming the high quality of the film. The device was fabricated in two lithographic steps (layers). In the first layer, e-beam lithography end e-beam evaporation was used to pattern two Ti (20 nm)/Au (30 nm) electrodes. A small gap (~50 nm) was left between both electrodes, so large electric fields could be generated by applying a few volts. The second layer consists on a $Al_2O_3$ (70 nm)/Ti (20 nm)/Au (30 nm) nanowire, patterned on top of the gap and running perpendicular to the first layer electrodes. E-beam lithography and e-beam evaporation was used for this purpose. Several of such devices are patterned in a single sapphire substrate. Optical lithography and reactive ion etching was used to remove the $VO_2$ outside of the gap area and isolate the different devices from each other. More information on the device fabrication process can be found in Extended Data Figure 1.

**Fast transport measurements.** Measurements were carried out in a TTPX Lakeshore cryogenic probe station. The station is equipped with high-speed (20 GHz) probes, with ground/line/ground geometry and 50 Ω characteristic impedance. In order to avoid reflections (the insulating resistance of the device is in the $10^4$ Ω range) a 50 Ω termination to ground was installed before the sample. A 240 MHz Tektronix function generator was used to create the voltage pulses and a 50 Ω terminated Tektronix broadband oscilloscope (20 GHZ) was used to monitor the current. The electrical circuit set up ensured a rise time around 5 ns.

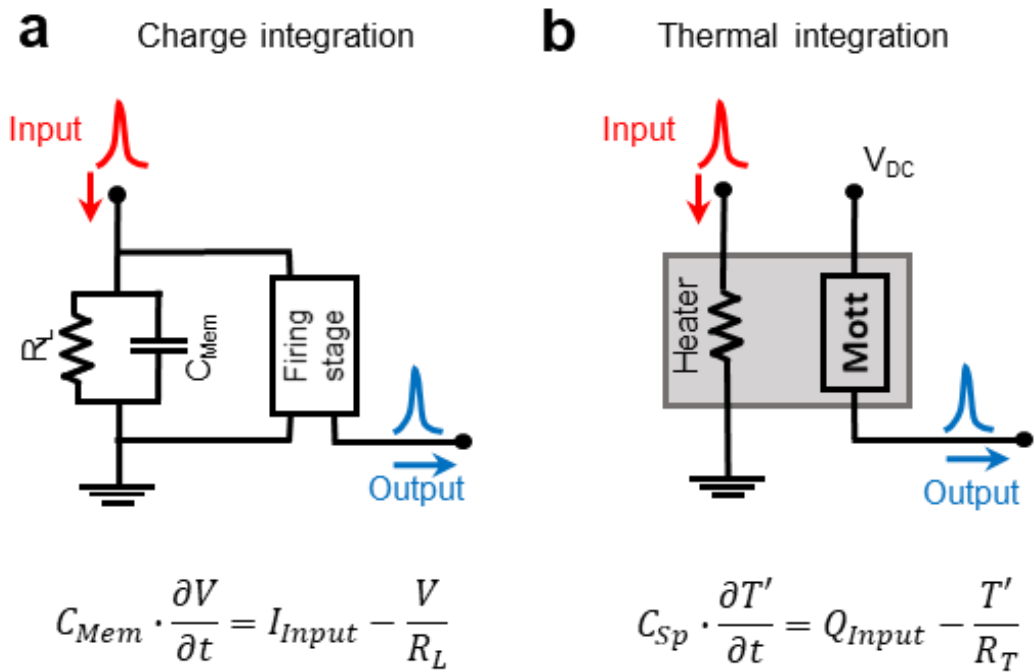

**Figure 1 | Charge vs thermal signal integration. a** Circuit representation of the leaky integrate-and-fire neuron model. Its dynamics are described by the equation shown below the circuit. A capacitor $C_{Mem}$ represents the neuron membrane capacitance, and accumulates charge from input current pulses $I_{Input}$ (red). The leaking resistor $R_L$ represents the leaky term of the equation. If the voltage $V$ across $C_{Mem}$ rises above a threshold, the firing stage will produce an output spike (blue). **b** Schematic representation of a thermal transfer-based neuristor. Incoming current pulses (red) dissipate power ($Q_{Input}$) in a heating resistance. A Mott nanodevice is kept under a DC voltage $V_{DC}$, and will emit a current spike (blue) if a threshold temperature is exceeded. The equation below describes the thermal dynamics of the system, where $C_{Sp}$ is the specific heat, $R_T$ the thermal resistance and $T' = T - T_{Eq}$: being $T$ the local temperature and $T_{Eq}$ the equilibrium (substrate) temperature.

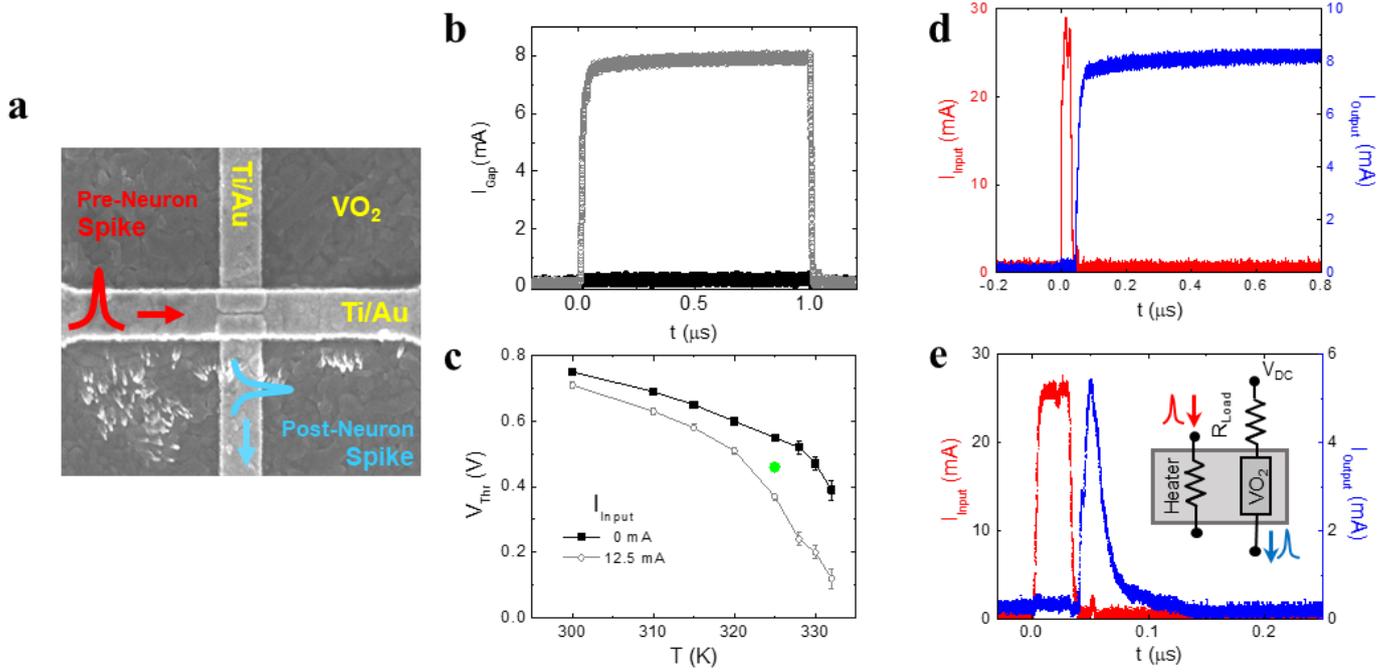

**Figure 2 | Experimental realization of the proposed neuristor. a** SEM image of the device. Two layers of electrodes, electrically isolated by an $Al_2O_3$ spacer, are visible. Bottom layer consist of two Ti/Au electrodes, running vertically in the image. A small gap (~50 nm) is left in between. The upper electrode is kept at $V_{DC}$ bias. The top electrode layer is a Ti/Au nanowire, running horizontally in the image. Input current pulses (red) are sent through that heating electrode, while the neuristor output (blue) is collected though the lower bottom electrode. **b** Current through a $VO_2$ gap vs time, when a 1 μs voltage pulse is applied. $T$=315 K. Two different pulse amplitudes are shown, 0.87 V (black) and 0.92 V (grey). **c** Threshold voltage vs temperature in a gap, when two different DC currents are applied to the heater: 0 mA (black) and 12.5 mA (grey). The green dot is an example of the $V_{DC}$ and temperature conditions for the device to behave as a neuristor. **d** Current vs time. Left axis (red curve) shows the input current through the heater. Right axis (blue curve) shows the output through the gap. Conditions were $V_{DC}$ = 0.85 $V_{Th}$ and $T$=325 K. **e** Current vs time, with a load resistor in series with the gap. Left axis (red) shows the input current pulse. Right axis (blue) shows the output through the gap. $R_{Load}$ = 10 kΩ, $V_{DC}$ = 0.88 $V_{Th}$ and $T$=325 K. Inset: schematic circuit showing $R_{Load}$ connected in series with the gap.

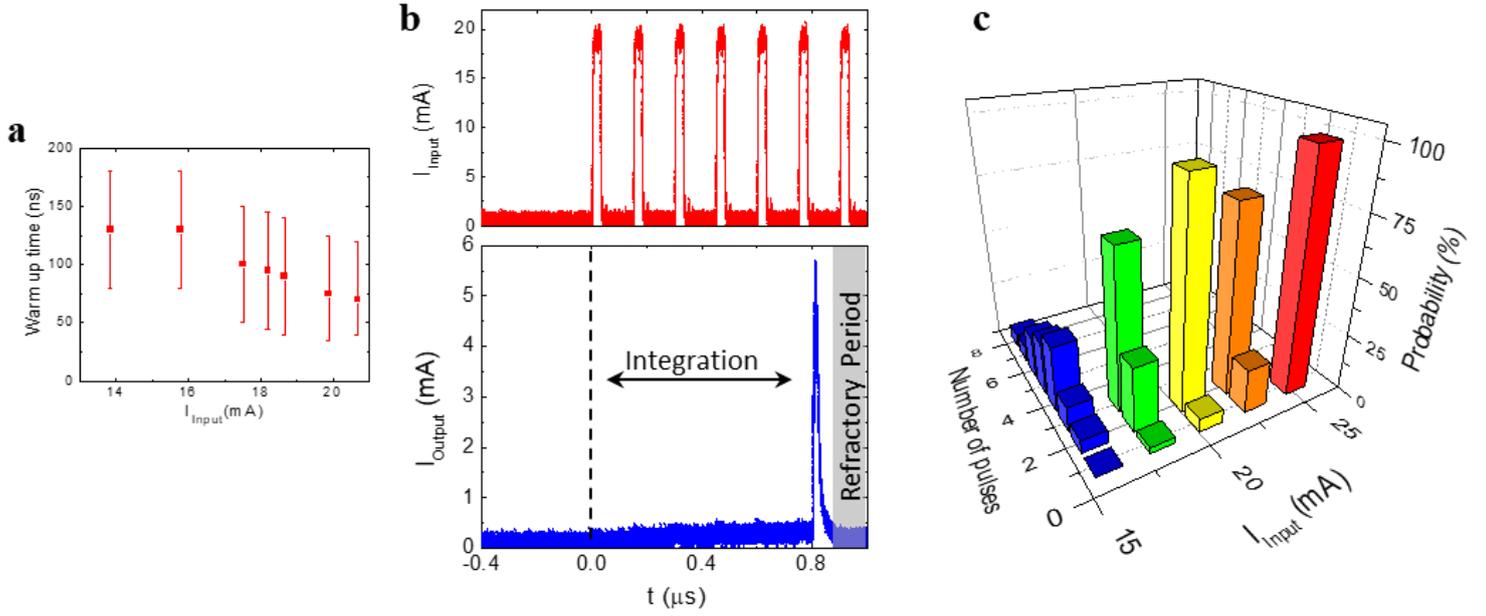

**Figure 3│ Leaky integrate and fire dynamics. a** Warm up time as a function of the input current. $V_{DC}$ = 0.85 $V_{Th}$ and $T$=325 K. **b** Current vs time when the neuristor is stimulated with a train of pulses. Upper panel shows the input current, consisting on 30 ns pulses with a 150 ns period. Bottom panel shows the output current through the gap. There is an integration time between the moment the pulses are applied and the moment the IMT is triggered. $R_{Load}$ = 10 kΩ, $V_{DC}$ = 0.88 $V_{Th}$ and $T$=325 K. No firing occurs after the last pulse because the neuron is in its refractory period (indicated with a grey shaded area). **c** Probability that the device will fire after a certain number of pulses. Pulses are 150 ns apart from each other. Several pulse amplitudes are shown. $V_{DC}$ = 0.91 $V_{Th}$ and $T$=325 K.

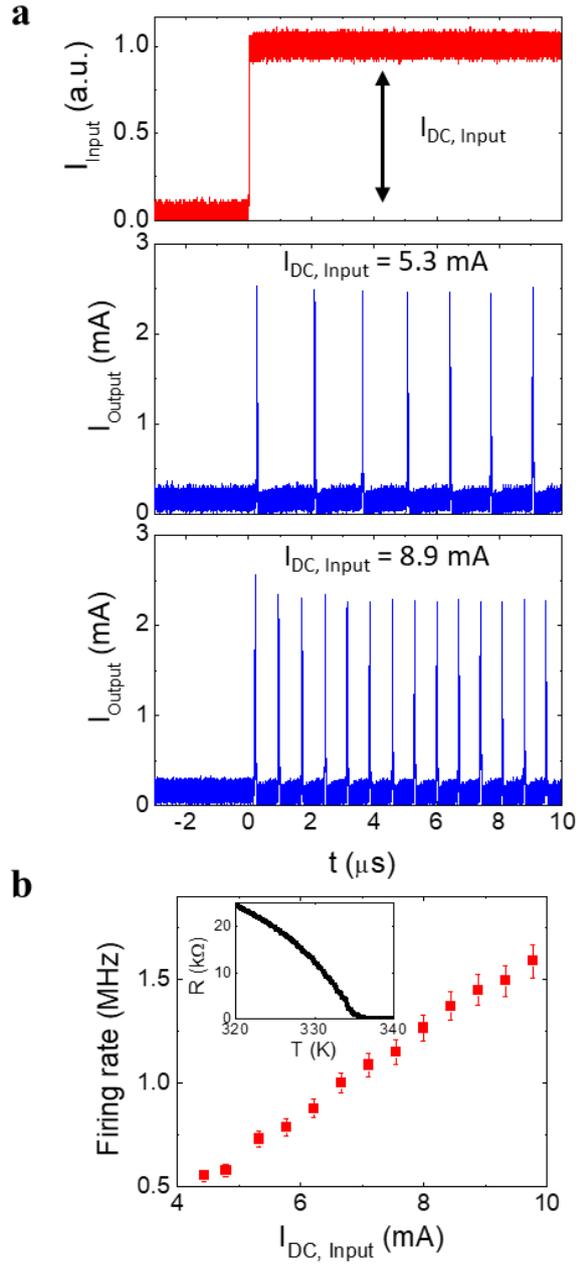

**Figure 4 | Rate coding. a** Current vs time when a DC current is applied as input. Top panel shows the input current $I_{DC,Input}$ through the heater (red). Middle and lower panels show the output spiking pattern (blue) induced in the gap. The responses to two $I_{DC,Input}$ values are shown: 5.3 mA (middle panel) and 8.9 mA (lower panel). $R_{Load}$ = 10 kΩ, $V_{DC}$ = 0.88 $V_{Th}$ and T=325 K. **b** Firing rate of the output spiking pattern as a function of $I_{DC,Input}$. Inset: resistance of a VO$_2$ gap as a function of the temperature. Only the cooling branch is shown.

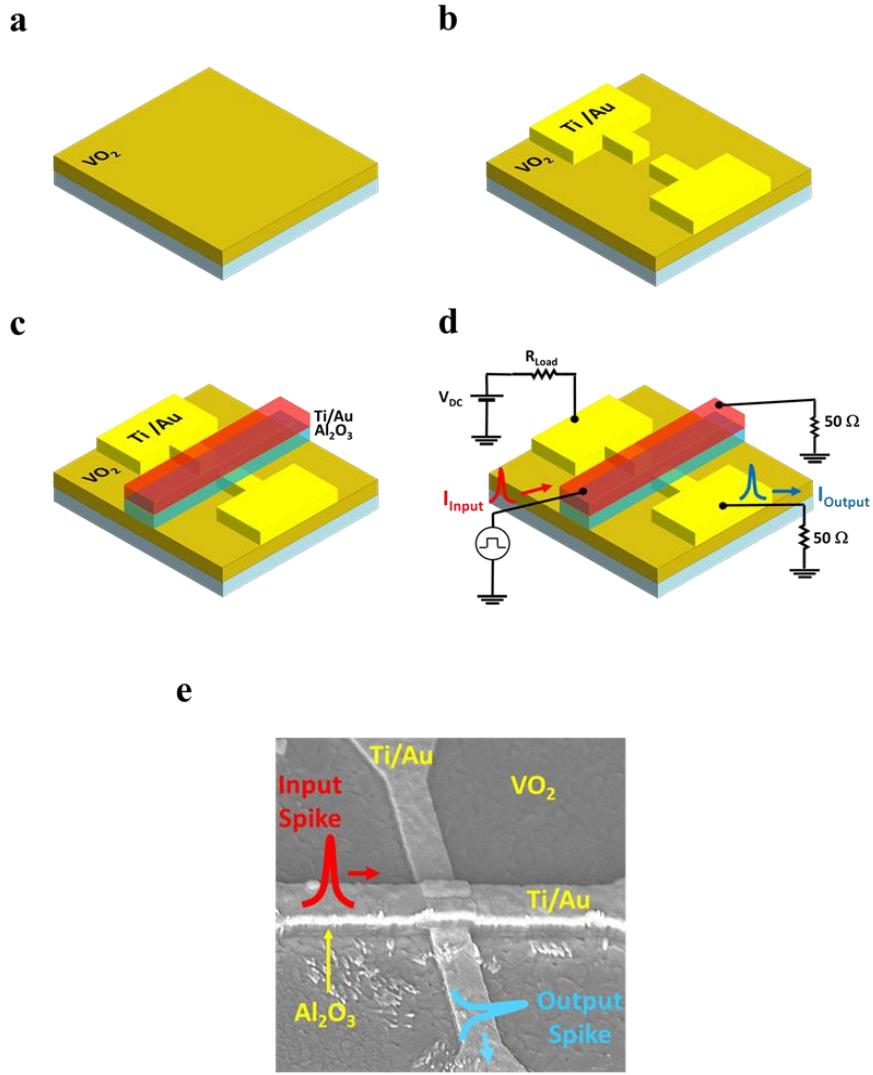

**Extended Data Figure 1 | Fabrication and configuration of the neuristor. a-d** different stages of the fabrication and set up process. **a** shows the VO$_2$ film (light brown) on top of a sapphire substrate (light blue). **b** shows the sample after a first lithographic step in which two Ti/Au electrodes (yellow) are patterned, leaving a small gap in between (~50 nm). **c** shows the sample after the second lithographic step in which the following trilayer was grown: Al$_2$O$_3$ (70 nm, blue) / Ti (20 nm, red) / Au (30 nm, red). **d** shows the electrical connections used for the fast transport measurements (R$_{Load}$ not included). **e** Tilted SEM image of the device. The Al$_2$O$_3$ and metallic layers of the heating electrode can be distinguished.